\documentclass[preprint,floats,aps,epsfig,nofootinbib,amssymb]{revtex4-1}
\usepackage{mathrsfs}

\usepackage{slashed}
\usepackage{graphicx,color}
\usepackage{dcolumn}
\usepackage{bm}
\usepackage{subfig}
\usepackage{graphicx}
\usepackage{amssymb}
\usepackage{stackrel}
\usepackage{pgfplots}
\usetikzlibrary{positioning}
\usetikzlibrary{snakes}

\usepackage{xcolor}
\usepackage[normalem]{ulem}
\usepackage{cancel}

\usepackage{soul}
\setstcolor{red}

\begin{document}

\title{CP violation in the top-assisted electroweak baryogenesis }

\author{Wei Chao$^1$}
\email{chaowei@bnu.edu.cn}
\author{Yandong Liu$^{2,3}$}
\email{liuyd@bnu.edu.cn}
\affiliation{$^1$Center for Advanced Quantum Studies, Department of Physics, Beijing Normal University, Beijing, 100875, China\\
$^2$Key Laboratory of Beam Technology of Ministry of Education,
College of Nuclear Science and Technology, Beijing Normal University, Beijing 100875, China \\
$^3$Beijing Radiation Center, Beijing 100875, China}
\vspace{3cm}

\begin{abstract}
The origin of the baryon asymmetry of the universe (BAU) is a longstanding problem in the high energy physics. The electroweak baryogenesis mechanism, which generates the BAU during the first order electroweak phase transition, provides a testable solution to this problem. In this paper we revisit the top-assisted electroweak baryogenesis model, which extends the standard model (SM) with only one scalar singlet. Constraints on the new CP-violating coupling as well as the mixing between the SM Higgs and the new scalar singlet are derived by considering the latest result of ACME and oblique observables. Furthermore, we derive  constraints on these two parameters arising from the data of Higgs measurement at the LHC as well as the four top-quark production at the LHC. 
\end{abstract}

\maketitle
\section{Introduction}

The standard model (SM) of particle physics  remarkably agrees with almost all high energy experimental results. However it can not be the fundamental theory, as there are solid evidences of new physics beyond the SM  arising from neutrino oscillations and astrophysical observations, of which the origin of the baryon asymmetry of the universe (BAU) is  a longstanding problem. Combing the result of the PLANK~\cite{Ade:2015xua} with that from the WMAP, the observed BAU is 
\begin{eqnarray}
Y\equiv {\rho_B \over s } = (8.61\pm 0.09) \times 10^{-11} \;, 
\end{eqnarray}
where $\rho_B$ is the baryon number density, $s$ is the entropy density of the universe. 
Assuming that our universe was matter-antimatter symmetric at the time of the Big Bang, there should be a mechanism that leads to the origin of the BAU during  the subsequent evolution. According to Sakharov~\cite{Sakharov:1967dj}, the following three criteria must be satisfied for a theory to generate the BAU: (1) baryon number violation; (2) C and CP violations; (3) a departure from the thermal equilibrium. The SM contains all these three ingredients. However the CP phase from the CKM mixing matrix  can not give rise to a large enough BAU, because QCD damping effects reduce the generated asymmetry to a negligible amount~\cite{Huet:1994jb}.

There are several successful baryogenesis mechanisms~\cite{Morrissey:2012db,Buchmuller:2005eh,Affleck:1984fy,Riotto:1999yt,Fukugita:1986hr}, of which the electroweak baryogenesis mechanism (EWBG)~\cite{Morrissey:2012db}, that generates the BAU during the strongly first order electroweak phase transition (EWPT),  is promising and attractive, as it can be tested in the energy, cosmic and intensity frontiers  by searching for new Higgs interactions and  the strength of EWPT at the Large Hadron Collider (LHC), detecting signals of stochastic gravitational wave~\cite{Caprini:2015zlo,Chao:2017vrq,Chao:2017ilw} arising from the first order EWPT in the  spaced based interferometer, and examining CP violations by measuring the electric dipole moments (EDMs) of electron,  neutron and molecular~\cite{Chupp:2017rkp}. 

In the EWBG, CP-violating (CPV) interactions on the bubble wall can lead to non-zero number densities of left-handed fermion and right-handed fermion with the same value but opposite sign, where the fermion may be the SM one  or brand new, such as neutralino or chargino in the MSSM.  Non-zero number densities can be diffused into the symmetric phase and be translated into non-zero number densities of other SM fermions via inelastic scatterings.  The sphaleron process will wash out number densities of left-handed fermions, resulting in a net BAU. Once eaten by the expanding bubble, this BAU will be kept since the sphaleron is decoupled inside the bubble.
In this paper we investigate the top-assisted EWBG mechanism~\cite{Espinosa:2011eu,Cline:2012hg,Chao:2017oux} by inspecting the collider signatures of this model at the LHC as well as the updated constraints from precision measurements.   We work in the framework of the SM extended with a real scalar singlet $\phi$, which is crucial for generating the strongly first order EWPT. The CPV top quark interaction is introduced with a dimension-5 effective operator: $\phi/\Lambda \overline{Q_{L}^{3} } \widetilde{H} (a + ib) t_R^{}  + {\rm h.c.}$~\cite{Espinosa:2011eu}, where $\Lambda $ is the cut-off scale, $Q_L^3$ is the third generation left-handed quark doublet, $t_R^{}$ is the right-handed top quark, $H$ is the SM Higgs doublet, $a$ and $b$ are dimensionless parameters. It should be mentioned that the CPV top interactions may also be given by a dimension-6 effective operator with a $Z_2$ discrete symmetry $\phi\to -\phi$,  in which $\phi$ serves as a cold dark matter candidate~\cite{Cline:2012hg,Cline:2017qpe,Chao:2015uoa}.   Our main results are listed as follows:
\begin{itemize}
\item The  CPV  coupling of the  SM Higgs with the top quark and the mixing angle between the SM Higgs and the scalar singlet are strongly constrained  by the latest ACME result of the electron EDM as well as oblique observables, see Fig.~\ref{fig:ST} for detail.

\item The production rate of the SM Higgs at the LHC depends sensitively on the CP property of the the top quark interaction \cite{Boudjema:2015nda}. The signal ratio yield to the standard model expectation puts a strong constraint on the CPV coupling, see  Fig.~\ref{fig:EDM} in the left-panel for detail.

\item  The four top quark production at the LHC, which is sensitive to the CP property of the top quark Yukawa interactions, puts strong constraint on the $C_\phi$, see the right-panel of the Fig.~\ref{fig:EDM} for detail. Combing all constraints together, we derive the updated constraints on this model as can be seen in  the Fig.~\ref{fig:fourtop}.  The available parameter space shrinks to a small regime near the origin at the $(0,~0)$. 

\end{itemize}
 
The remaining of the paper is organized as follows: In section II we present the basic setup of the model. Section III is focused on various constraints on the parameter space of the model. In section IV we discuss the collider signatures of the model at the LHC.  The last part is concluding remarks.

\section{The model}

In this section, we present the detail of the model. We work in the framework of the minimal SM extended with a real scalar singlet $\phi$. The model is remarkably simple and was studied in many references~\cite{Burgess:2000yq,McDonald:1993ex,Barger:2007im}.  The most general renormalizable Higgs potential can be written as
\begin{eqnarray}
V&=& -\mu^2 H^\dagger H + \lambda (H^\dagger H)^2 - {1\over 2 }\mu_\phi^2 \phi^2 + {1\over 4} \lambda_1 \phi^4 + \lambda_2^{} \phi^2 (H^\dagger H) \nonumber \\ && + \rho^3 \phi + {1\over 3} \Lambda_1 \phi^3 + \Lambda_2^{} \phi(H^\dagger H) \label{potential}
\end{eqnarray}
where $H$ is the SM Higgs doublet, $\mu, \mu_\phi^{}, \rho^{}, \Lambda_1^{}, \Lambda_2^{}$ have dimensions of mass, $\lambda$, $\lambda_1$ and $\lambda_2$ are dimensionless couplings, the first five terms in the potential respect a $Z_2$ symmetry for the $\phi$ field, while the final three terms do not. We will not include the linear term, $\rho^3 s$ in the following study, since one can always remove this term by shifting $\phi\to \phi-\delta$. As a result, there are seven free parameters in the potential. 
Taking $v_h$ and $v_\phi$ as the vacuum expectation values (VEVs) of the SM Higgs and $\phi$ at the zero temperature, one can use the tad-pole conditions to express $\mu^2$ and $\mu_\phi^2$ in terms of physical parameters $v_h^{}$ and $v_\phi^{}$,
\begin{eqnarray}
&&-\mu^2 + \lambda v_h^2  + \lambda_2^{}  v_\phi^2 + \Lambda_2^{}  v_\phi^{} =0 \\
&&-\mu_\phi^2 + \lambda_1^{}  v_\phi^2 + \lambda_2^{}  v_h^2 + \Lambda_1^{}  v_\phi^{}  + {1\over 2 }\Lambda_2^{}   {v_h^2\over v_\phi^{} } =0 
\end{eqnarray}
The mass terms of neutral scalars can be written as
\begin{eqnarray}
{\cal L}_{\rm mass}^{} ={1\over 2 } \left( \matrix{h& \phi}\right)\left( \matrix{ 2 \lambda v_h^2  & v_h^{}  \Lambda_2^{} + 2 \lambda_2^{} v_h^{} v_\phi^{} \cr  v_h^{}  \Lambda_2^{} + 2 \lambda_2^{} v_h^{} v_\phi^{}  
& 2 \lambda_1^{} v_\phi^2 +\Lambda_1^{} v_\phi^{} -  {1\over 2}\Lambda_2^{} v_h^2 v_\phi^{-1}}\right)\left( \matrix{h\cr \phi} \right)
\end{eqnarray}
where the mass matrix can be diagonalized by a $2 \times 2$ unitary transformation. $\lambda, \lambda_1$ and $\lambda_2$ can be reconstructed by the mass eigenvalues $m_h^2$, $m_\phi^2$, the mixing angle $\theta$ between $h$ and $\phi$, $v_h$, $v_s$, $\Lambda_1^{} $ and $\Lambda_2^{} $, which are  physical parameters of the potential.   

The Yukawa interactions of  the top-quark can be written as 
\begin{eqnarray}
y_t^{} \overline{Q_L^3} \widetilde{H}   t_R^{} + {\zeta \Lambda^{-1}} \overline{Q_L^3} \widetilde{H} \phi  t_R^{} + {
\rm h.c. } \label{dimension5}
\end{eqnarray}
where $y_t$ is the SM top Yukawa coupling, $\Lambda$ has mass dimension and serves as the cut-off scale, $Q_L^3$ is the third generation quark doublet. As was shown in Ref.~\cite{Chao:2017oux},  this high dimensional effective operator may come from integrating out a vector-like top quark.  In the Eq.(\ref{dimension5}), there is a rephasing invariant ${\rm Arg} [\zeta y_t^{}]$ that can not be rotated away by fields redefinition, resulting in a CPV phase. By setting $\zeta=a+ib$, the  Eq. (\ref{dimension5}) can be written as
\begin{eqnarray}
{1 \over \sqrt{2} } \left( y_t^{}  v_h^{} + a \delta    v_h\right) \bar t t  + {1\over \sqrt{2}} b \delta  v_h \bar t i\gamma_5 t + {1\over \sqrt{2}} (y_t + a \delta)  h \bar t t + {1\over \sqrt{2}} \delta b h \bar t i\gamma_5 t  \nonumber \\+ {1\over \sqrt{2}} {v_h \over\Lambda } a \phi \bar t t +  {1\over \sqrt{2}} {v_h \over\Lambda } b \phi \bar t i\gamma_5 t + {1\over \sqrt{2}} {1\over \Lambda} a h \phi \bar t t + {1\over \sqrt{2}} {1\over \Lambda} b h \phi \bar t i\gamma_5 t  \label{massterm}
\end{eqnarray}
where $\delta \equiv v_\phi /\Lambda$. The first two terms are the mass term of the top quark.  If we ignore the mixing of the top quark with light quarks, which is tiny as predicted by the experimental value of the CKM matrix, it is necessary to perform a chiral rotation  to have a defined field with a real mass term
\begin{eqnarray}
t\to \exp(i \alpha \gamma_5^{} ) t \label{rotation}
\end{eqnarray}
where $\alpha$ is a real parameter. Taking Eq.~(\ref{rotation}) into Eq. (\ref{massterm}) and eliminating the parity-violating mass term, one arrives at
\begin{eqnarray}
\tan 2 \alpha = - { b\delta \over y_t^{} + a \delta }, \hspace{0.5cm}  m_t \equiv {v_h Y_t^{}  \over \sqrt{2}}= {v_h \over \sqrt{2} }\sqrt{ \left(y_t +a\delta \right)^2 + {b^2 \delta^2}} \; ,
\end{eqnarray}
where $Y_t$ is defined as the effective Yukawa coupling. 
We further define $ h = c_\theta \hat h - s_\theta \hat \phi$, $\phi = c_\theta \hat \phi + s_\theta \hat h$, where $c_\theta=\cos \theta$, $s_\theta=\sin \theta$, $\hat h$ and $\hat \phi$ are the mass eigenstates.  The top quark Yukawa interactions can be rewritten as 
\begin{eqnarray}
{1\over \sqrt{2}}\bar{t}(S_\phi s_\theta +Y_t c_\theta+ i \gamma^5 C_\phi s_\theta)t \hat{h}  \label{hinter}  \\
{1\over \sqrt{2}}\bar{t}( S_\phi c_\theta- Y_t s_\theta + i \gamma^5 C_\phi c_\theta)t \hat{\phi} \label{phiinter}
\end{eqnarray}
where 
\begin{eqnarray}
S_\phi &=&{ Y_t^2 -y_t^2 - a \delta y_t^{}  \over Y_t} {v_h\over v_\phi} \\
C_\phi &=&{y_t b \delta \over Y_t } {v_h\over v_\phi} 
\end{eqnarray}
Interactions of  $\hat{h}$ with other SM particles is rescaled by a factor of $c_\theta$.

\section{Constraints}
In this section we study constraints on the parameter space of the model. We first discuss constraints of the EWBG. According to Sakharov, strongly first order EWPT is essential for generating the BAU during the electroweak symmetry breaking.  The SM Higgs itself is too heavy to saturate a first order EWPT, so extensions to the minimal Higgs sector is necessary. For an extended scalar sector with $Z_2$ symmetry the barrier between the symmetric phase and the broken phase, as required by the first order EWPT, arises from radiative corrections, which suffers from the gauge-dependence problem~\cite{Patel:2011th}. Although several attempts are made to  address this problem, there is no promising solution.  This problem can be avoided in our model as the barrier may arise at the tree level, while the gauge-dependent terms are sub-dominate and can be safely neglected.  The criterion for the strongly first order EWPT, $v_h/T_C>1$, required by quenching the sphaleron process inside the bubble, can easily be met for a large range of  scalar singlet mass~\cite{Barger:2007im}.  

Another constraint is from the bubble wall velocity.  A dedicate calculation given in Ref.~\cite{Kozaczuk:2015owa} shows that bubble wall velocity in the singlet driven EWBG with tree-level cubic term is large but still compatible with the EWBG. Alternatively a large bubble wall velocity will enhance the stochastic gravitational wave signal emitted during the EWPT, which can be tested in future space based interferometer such as LISA, Taiji etc~\cite{Chao:2017vrq}. It should be mentioned that further investigation with more careful treatment to the friction term in this model is still needed. Since the connection between strongly first order EWPT and the CPV Higgs interaction is weak and the constraint from the EWBG highly depends on the bubble dynamic, we will not consider their constraints in this paper.

\subsection{Oblique observables}
We study the updated constraints on the parameter parameter space of the model arising from electroweak precision measurements, i.e., oblique observables~\cite{Peskin:1991sw,Peskin:1990zt}, which are defined in terms of contributions to the vacuum polarizations of gauge bosons. In our model, $\phi$ contributes to the oblique parameter via its mixing with the SM Higgs. Thus $\delta S$ and $\delta T$ are proportional to $s^2_\theta$, and can be approximately written as
\begin{eqnarray}
\delta S &=& {s_\theta^2 \over 24 \pi } \left\{ \log R_{\phi h}^{} + \hat G(m_\phi^2, m_Z^2 ) -\hat G (m_h^2, m_Z^2)  \right \} \label{deltas} \\
\delta T &=& {3 s_\theta^2 \over 16\pi s_W^2 m_W^2 } \left[ m_Z^2 \left( {\log R_{Z \phi} \over 1-R_{Z \phi}^{} } - {\log R_{Z h} \over 1-R_{Z h }^{} } \right) -m_W^2 \left( {\log R_{W \phi} \over 1-R_{W \phi}^{} } - {\log R_{W h} \over 1-R_{W h }^{} } \right)\right] \label{deltat}
\end{eqnarray}  
where $s_W=\sin \theta_W$ with $\theta_W$ the weak mixing angle, $R_{ij}=m_i^2 /m_j^2$ and 
\begin{eqnarray}
\hat G (m_i^2, m_j^2 ) &=& - {79\over 3} + 9 R_{ij}^{} -2 R_{ij}^2 + (12 -4 R_{ij}^{} + R_{ij}^2 ) \hat F(R_{ij}) \nonumber \\
&& + \left(-10 +18R_{ij}^{} -6 R_{ij}^2 +R_{ij}^3 + 9 {1 + R_{ij} \over 1-R_{ij}}\right) \log R_{ij}^{}
\end{eqnarray}
with the expression of $\hat F(R_{ij})$ given in Ref.~\cite{Farzinnia:2013pga}.  

\begin{figure}[t]
\begin{center}
\includegraphics[width=0.435\textwidth]{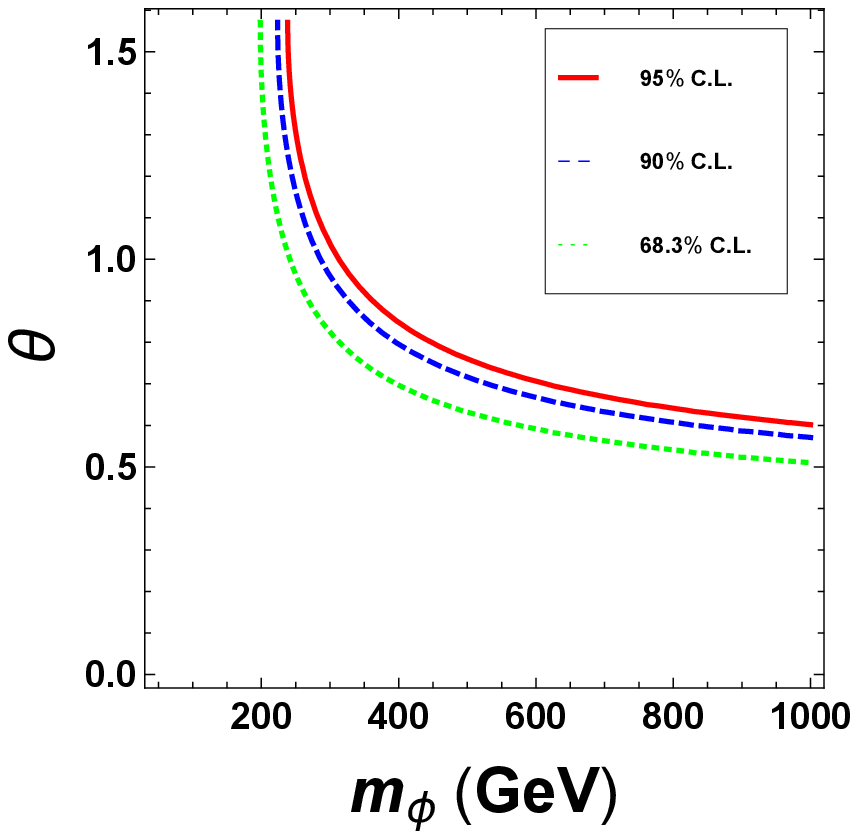}
\includegraphics[width=0.45\textwidth]{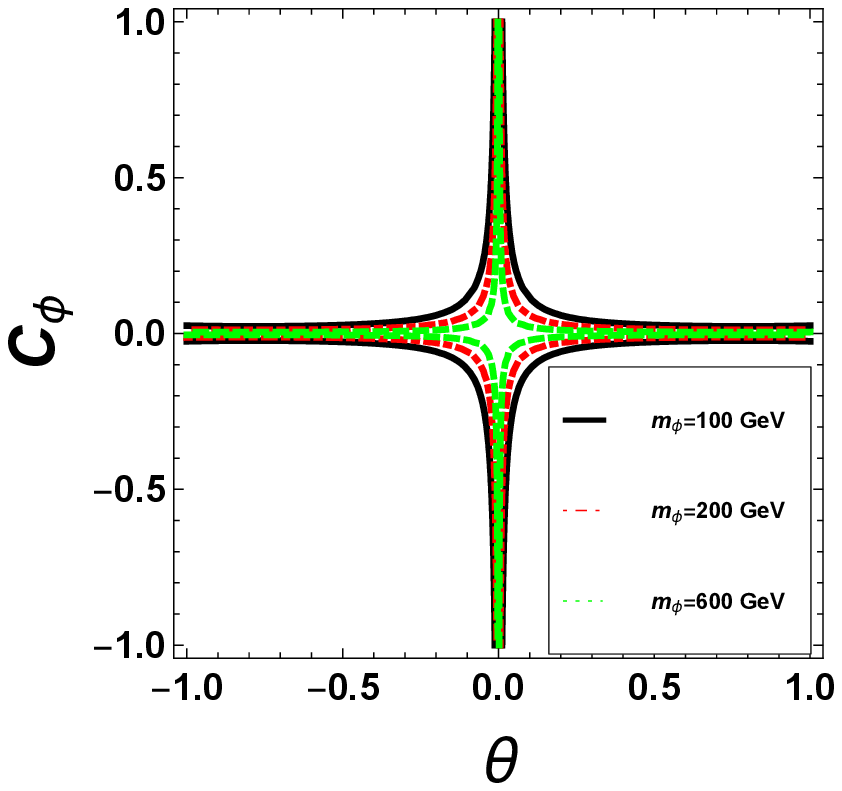}
\end{center}
\caption{ Left-panel: constraints from the oblique parameters. The sold, dashed and dotted lines correspond to constraints at the 95\% C.L., 90\% C.L. and 68.3\% C.L., respectively. The region to the above of various cure is excluded.  Right-panel:  Region plot of the electron EDM in the $\theta-C_\phi$ plane. The solid, dot-dashed and dashed lines correspond to $m_{\phi}=100$ GeV, 200 GeV and 600 GeV, respectively.}
\label{fig:ST}
\end{figure}

Notice that there are two free parameters, $\theta$ and $m_\phi^{}$, in Eqs.~(\ref{deltas}) and (\ref{deltat}). They are thus constrained by the oblique observables. 
There are many studies focusing on this issue,  but a revisiting to the same problem with updated global fitting results still make sense.  Using updated  reference values $m_{H, {\rm ref}}^{} =125~{\rm GeV}$ and $m_{t,{\rm ref}}=172.5~{\rm GeV}$, it has $S|_{U=0}=0.04\pm0.08$ and $T|_{U=0}=0.08\pm0.07$~\cite{Haller:2018nnx}, with the correlation coefficient $+0.92$. Constraints can be derived by performing $\delta \chi^2 $ fit to the data given above, with 
\begin{eqnarray}
\delta \chi^2 = \sum_{ij}^2 (\delta {\cal O}_i -\delta {\cal O}_i^0 ) (\sigma^2_{ij})^{-1} (\delta {\cal O}_j^{} - \delta {\cal O}_j^0 )
\end{eqnarray}
where $\sigma^2_{ij} =\sigma_i^{} \rho_{ij}^{} \sigma_j^{}$.

We show in the left-panel of the Fig.~\ref{fig:ST} the constraint of oblique parameters in the $m_\phi^{}-\theta$ plane. The sold, dashed and dotted lines correspond to constraints at the 95\% C.L., 90\% C.L. and 68.3\% C.L., respectively. The  region to the above of these cures are excluded.  For a fixed mass of the scalar singlet there is an upper bound on the mixing angle. 

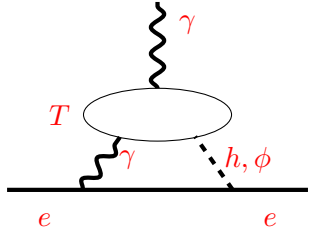
\begin{figure}[t]
\begin{center}
\begin{tikzpicture}
\draw[-,ultra thick] (-1,0)--(3,0);
\draw [-,snake=snake, ultra thick] (0,0) -- (0.5,0.7);
\draw [-,dashed, ultra thick] (2, 0) -- (1.5, 0.71);
\draw [-,snake=snake, ultra thick] (1,1.35) -- (1,2.5);
\draw (1,1) ellipse (28pt and 10 pt);
\node[red, thick] at (-0.5,-0.4) {$e$};
\node[red, thick] at (2.5,-0.4) {$e$};
\node[red, thick] at (-0.3,1.0) {$T$};
\node[red, thick] at (1.4,2.2) {$\gamma$};
\node[red, thick] at (0.6,0.4) {$\gamma$};
\node[red, thick] at (2.2,0.4) {$h,\phi$};
\end{tikzpicture}
    \caption{Two-loop Feynman diagrams for  the electron EDM.}\label{twoloop}
\end{center}
\end{figure}

\subsection{The electron EDM}
We now study constraint on the model from the electron electric dipole moment (EDM) measurement.  The CPV Yukawa interactions in Eqs.(\ref{hinter}) and (\ref{phiinter}) induce the EDM for the electron, which is dominated by the two-loop Barr-Zee diagram~\cite{Barr:1990vd}. The relevant Feynman diagram is given in the Fig.~\ref{twoloop}, where the contribution arises from exchange of a neutral scalar and a photon. Specializing the well-known result to our case, we arrive at the following equation for the electron EDM~\cite{Weinberg:1990me}
\begin{eqnarray}
d_e =\sqrt{2} d_e^{(2l)}c_\theta^{} s_\theta^{} C_\phi^{} {  v_h^{} \over m_t} \left[   g\left({m_t^2 \over m_h^2}\right) - g\left( {m_t^2\over m_\phi^2}  \right)\right]
\end{eqnarray}   
where $m_t$ is the top-quark mass, $d_e^{(2l)} \approx 2.5\times 10^{-27}$ $e\cdot cm$ quantifying  the two-loop benchmark EDM scale, the loop function is given by
\begin{eqnarray}
g(x) ={x\over 2 } \int_0^{1}dt {1\over t(1-t) -x } \ln \left[ {t(1-t) \over x} \right]\; .
\end{eqnarray}
 One has $g(x)\sim {1\over 2 } \ln x$ for  large $x$.

The updated constraint on the electron EDM  is~\cite{Andreev:2018ayy}
\begin{eqnarray}
|d_e^{}| <1.1\times10^{-29} ~{e\cdot cm} \; ,
\end{eqnarray}
which is given by the ACME collaboration that use THO molecules to constrain the electron EDM.  As an illustration, we show in the right panel of the Fig.~\ref{fig:ST} the region  allowed by the current ACME result in the $\theta-C_\phi$ plane by setting  $m_h=125~{\rm GeV}$ and $m_t=172.9~{\rm GeV}$. The black solid line, red dot-dashed line and green dotted line correspond to the case of $m_\phi=100~{\rm GeV}$, $200~{\rm GeV}$, and $600~{\rm GeV}$, respectively,  the region to the outside of these lines is excluded.

\begin{figure}[t]
\begin{center}
\includegraphics[width=0.45\textwidth]{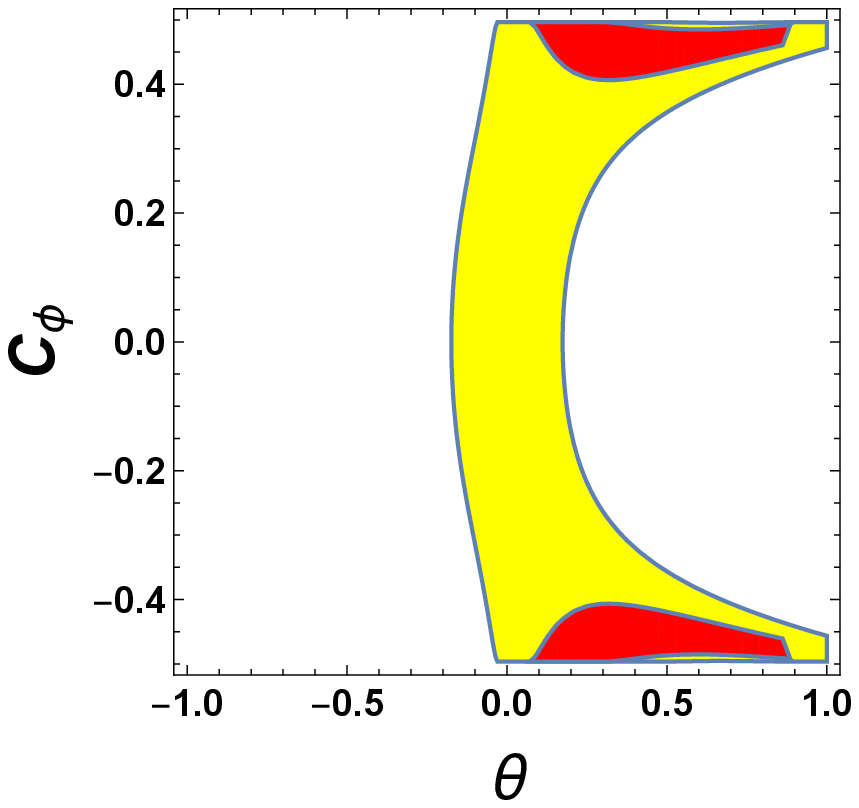}
\includegraphics[width=0.45\textwidth]{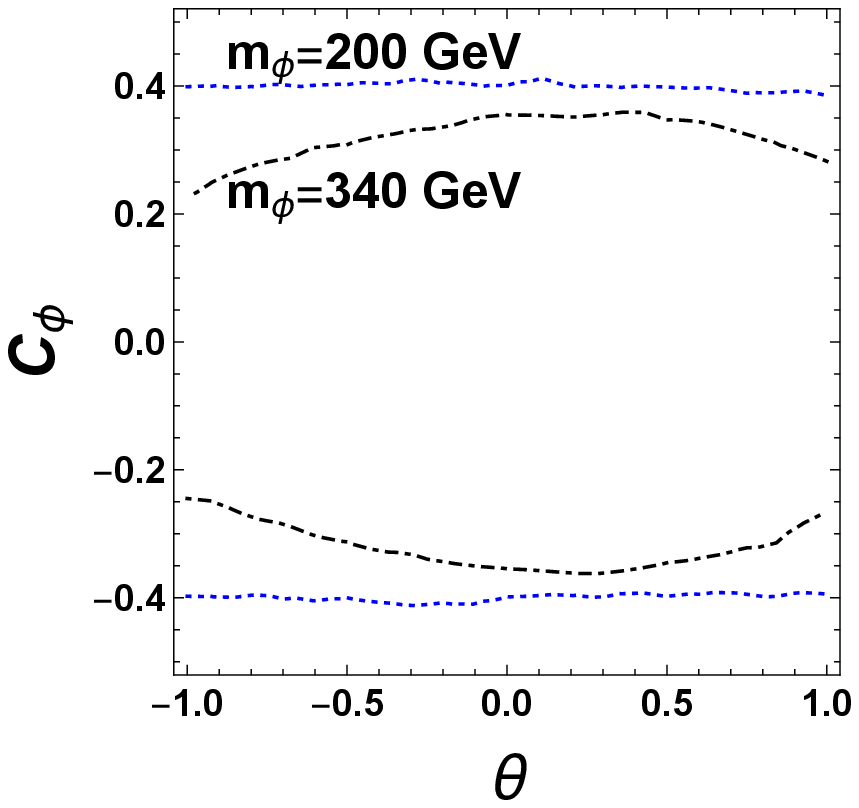}
\end{center}
\caption{ Left-panel: Constraints arising from Higgs measurement at the LHC in the  $\theta-C_\phi$ plane with the red and yellow regions correspond to the fit of the ratio at the $1\sigma$ and $2\sigma$ level, respectively. Right-panel: Constraint arising from four top quark production at the LHC in the $\theta-C_\phi$ plane at the $2\sigma$ confidence level.}
\label{fig:EDM}
\end{figure}

\section{ Higgs boson and multiple top-quark production at the LHC}
As can be seen in the section II, the CPV interaction between the SM Higgs and the top quark is induced through its mixing with $\phi$. In addition to  the CPV interaction, the mixing will also induce a universal suppression factor $\sim c_\theta$ in the couplings of the SM Higgs with the SM particles except for that of the top quark, which is  already given in the Eq.~(\ref{hinter}). 
They  will affect the single Higgs and multiple top  productions at the LHC,  as the single Higgs is produced at the LHC via the gluon fusion, which is dominated by the Yukawa interaction of the top quark and the electroweak contribution in the four top quark production at the LHC is comparable to QCD contribution~\cite{Alwall:2014hca,Frederix:2017wme}. 
%
Since the pair production of top-quark at the LHC is mainly induced by QCD interactions and the effect of top-Higgs interaction is only sub-dominate, there is almost no constraint from the top quark pair production  process.

\begin{figure}
\centering
\includegraphics[width=0.6\textwidth]{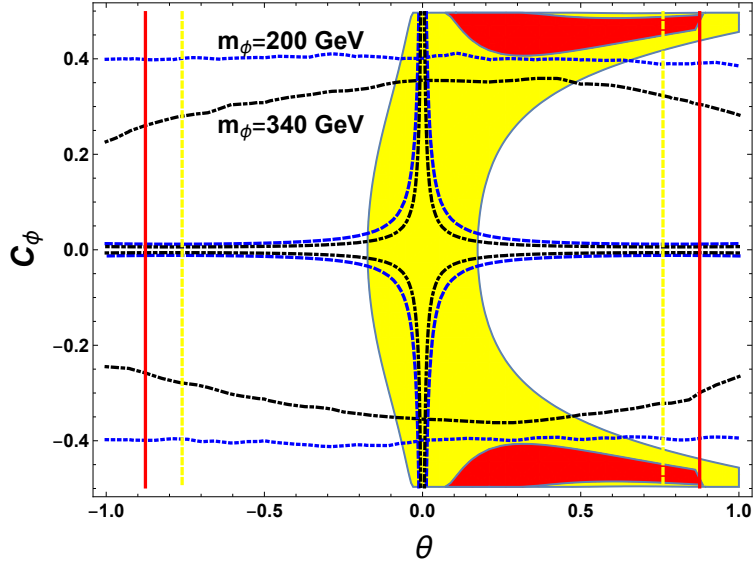}
\caption{\label{fig:fourtop} Combined constraints in the in the $\theta-C_\phi$ plane. The black dot-dashed  and blue dashed lines correspond to $m_\phi=340~{\rm GeV}$ and $200~{\rm GeV}$, respectively. The  red solid and  yellow dashed vertical lines are constraints arising from the oblique observables at the 68.3\% and 95\% C.L. respectively, by setting $m_\phi=340~{\rm GeV}$.}
\end{figure}

The top-Higgs interaction has been found indirectly after the Higgs boson discovery~\cite{Aaboud:2018urx,Sirunyan:2018hoz}. The next task is to measure its  CP property. The Higgs production rate highly depend on the CP property of the top-Higgs interaction and the production ratio with respect to the SM expectation can be written as
\begin{eqnarray}
\mu_{\rm pro}^{} =\frac{\sigma(gg\to H)}{\sigma(gg\to H)_{\text{SM}}} = \left(\frac{S_\phi s_\theta +Y_t c_\theta}{Y_t}\right)^2 + 2.26 \left(\frac{C_\phi s_\theta}{Y_t} \right)^2 \; . \label{ratiox}
\end{eqnarray} 
Obviously the contribution of the  CPV interaction  is enhanced by a factor of $2.26$ compared to the contribution of the CP conserving interactions.  as can be seen in the second term on the right-handed side of the eq.~(\ref{ratiox}).
Although the decay rates of the SM Higgs to $\gamma \gamma$, $\ell^+\ell^-$, $WW$, $ZZ$, $\bar bb$,   are rescaled by a factor  of $c_\theta^2$, its branching ratio to each channel  is unchanged as the decay channel is not changed. As a result, the signal ratio associated with Higgs measurements with respect to the standard model expectation is exactly the same as $\mu_{\rm pro}$.   The ATLAS and CMS~\cite{Sirunyan:2018koj} collaborations have measured the production and the decay of the SM Higgs at the LHC, and the best fit of the signal ratio is $\mu=1.17\pm0.10$, which alternatively put constraint on $\theta$ and $C_\phi$.
As an illustration, we show in the left-panel of the Fig.~\ref{fig:EDM}, constraints arising from the Higgs measurements in the $\theta-C_\phi$ plane, where the red and yellow regions are allowed at the $1\sigma$ and $2\sigma$ confidence level, respectively.  We have assumed that $a=0$, $v_\phi=v_h$, and $m_t=172.9~{\rm GeV}$ when making the plot.
This constraint, combined with that of the EDM and four top quark production at the LHC, will yield a strong constraint on the parameter space as can be seen in the Fig.~\ref{fig:fourtop}.

It is found that the four top quark production~\cite{Cao:2019ygh} severely depends on the CP property of the top quark Yukawa interaction, namely it is destructive with the gauge bosons ($g/Z/\gamma$) mediation for the CP even interaction while constructive for the CP odd interaction. In the alignment limit $\theta=0$ and scalar mass degeneracy limit $m_{\hat{h}}=m_{\hat{\phi}}$,
the four top quark production cross section, calculated utilizing the MadEvent, can be written as
\begin{eqnarray}
\sigma(\bar{t} t \bar{t}t)_{\text{13~TeV}}&=&9.998-1.522\frac{ Y_t^2 + C_\phi^2 }{Y_t^2}+2.883\left(\frac{C_\phi}{Y_t}\right)^2 
+ 1.173\frac{\left( Y_t^2 + C_\phi^2  \right)^2}{Y_t^4}  \nonumber \\
&+& 2.713\frac{ Y_t^2 + C_\phi^2 }{Y_t^2}\left(\frac{C_\phi}{Y_t}\right)^2 +1.827\left(\frac{C_\phi}{Y_t}\right)^4,
\end{eqnarray}
which clearly shows the inference effect depends on the CP property of Yukawa interaction.

The experimental collaboration has searched the four top quark productions at the 13 TeV LHC with the integrated luminosity of 137 fb$^{-1}$~\cite{Sirunyan:2019wxt}.  The observed significance of the SM predicted process is about $2.6\sigma$.
The upper limit on the $t\bar{t}t\bar{t}$ cross section is $8.5^{+3.9}_{-2.6}$ fb with the assumption of no SM contribution in terms of BDT methods. Thus  the upper limit on the four top quark production cross section is 16.3 fb at the $95\%$ C.L..

As an illustration, we show in the right-panel of the Fig. \ref{fig:EDM}, contours of  the four top quark production cross section in the $\theta-C_\phi$ plane.  As $ m_\phi \sim200$ GeV, $\hat \phi$ has the almost same contribution to the production cross section as the SM Higgs, which means the four top quark production cross section mildly depends on the mixing angle $\theta$. When $m_\phi\sim 340$ GeV which is near the top quark pair threshold, it has larger contribution than that from SM-like Higgs boson $\hat{h}$, namely the contour line shrinks towards the small $\theta$ region. We show in the Fig.~\ref{fig:fourtop}, the combined constraints on this model. The blue dashed and black dot-dashed lines correspond to $m_\phi=200~{\rm GeV}$ and $340$ GeV, respectively.  The red solid and yellow dashed vertical lines are the constraint of oblique observables at the 68.3\% and 95\% C.L., by setting $m_\phi=340~{\rm GeV}$.   The constraint of Higgs measurement does not depend on the mass of the singlet.  It shows that both $C_\phi$ and $\theta$ are stongly constrained and the available parameter space shrinks to the  regime near the origin at $(0,~0)$. 


\section{conclusions}

The electroweak bayogenesis mechanism is an attractive solution the BAU due to its testability. In this paper we hunt for the CP violation in the top-assisted electroweak baryogensis, which is one of the most economic extensions to the SM and contains only a few new parameters.  New constraints from precision observables, EDM measurement, Higgs measurement at the LHC and four top quark production at the LHC were derived.  The available parameter space is shrunk  to a small region near the origin (0,~0) in the $C_\phi-\theta$ plane. Once the mixing angle $\theta$ is determined in the future, this study will provide a solid guidance to the search of  CP violating top interaction. It should be mentioned that constraints on the model from the  EWBG is not included as there are large uncertainties in the calculation of bubble wall velocity, which, although interesting but beyond the reach of this paper, will be presented in a future study.

\begin{acknowledgments}
This work was supported by the National Natural Science Foundation of China under grant No. 11775025, No. 11805013 and the Fundamental Research Funds for the Central Universities under grant No. 2017NT17, No. 2018NTST09.
\end{acknowledgments}

\end{document}